\documentclass[10pt,twocolumn,aps,pre,superscriptaddress]{revtex4-1} %

\usepackage{amsmath}
\usepackage{graphicx}
\usepackage{color}
\usepackage{upgreek}
\usepackage[linkcolor = blue, citecolor = blue, urlcolor = blue, colorlinks = true]{hyperref}
\usepackage[version=3]{mhchem}
\usepackage{commath,amssymb}
\usepackage{ushort}

\usepackage{bm}
\usepackage[capitalise]{cleveref}
\usepackage{textcomp}

\graphicspath{{./figures/}}

\newcommand{\second}{\, \mathrm{s}}

\usepackage[exponent-product = \times]{siunitx}

\newtoks\rowvectoks
\newcommand{\rowvec}[2]{%
 \rowvectoks={#2}\count255=#1\relax
 \advance\count255 by -1
 \rowvecnexta}
\newcommand{\rowvecnexta}{%
 \ifnum\count255>0
 \expandafter\rowvecnextb
 \else
 \begin{pmatrix}\the\rowvectoks\end{pmatrix}
 \fi}
\newcommand\rowvecnextb[1]{%
 \rowvectoks=\expandafter{\the\rowvectoks&#1}%
 \advance\count255 by -1
 \rowvecnexta}

\renewcommand{\vec}[1]{\mathbf{#1}}
\newcommand{\acos}{\mathrm{acos}}

\begin{document}
\author{Patrick Kreissl}
\affiliation{Institute for Computational Physics (ICP), University of Stuttgart, Allmandring 3, 70569 Stuttgart, Germany}

\author{Christian Holm}
\affiliation{Institute for Computational Physics (ICP), University of Stuttgart, Allmandring 3, 70569 Stuttgart, Germany}

\author{Joost de Graaf}
\email{jgraaf@icp.uni-stuttgart.de}
\affiliation{Institute for Computational Physics (ICP), University of Stuttgart, Allmandring 3, 70569 Stuttgart, Germany}

\title{The Efficiency of Self-Phoretic Propulsion Mechanisms with Surface Reaction Heterogeneity}

\date{\today}

\begin{abstract}
We consider the efficiency of self-phoretic colloidal particles (swimmers) as a function of the heterogeneity in the surface reaction rate. The set of fluid, species, and electrostatic continuity equations is solved analytically using a linearization and numerically using a finite-element method. To compare spherical swimmers of different size and with heterogeneous catalytic conversion rates, a `swimmer efficiency' functional~$\eta$ is introduced. It is proven, that in order to obtain maximum swimmer efficiency the reactivity has to be localized at the pole(s). Our results also shed light on the sensitivity of the propulsion speed to details of the surface reactivity, a property that is notoriously hard to measure. This insight can be utilized in the design of new self-phoretic swimmers.
\end{abstract}

\maketitle

\section{Introduction\label{sec:intro}}

In recent years there has been a surge of interest in the study of active matter~\cite{ramaswamy10a,marchetti13a}, in particular concerning self-propelled colloidal particles (swimmers)~\cite{ebbens10a,hong10a,sengupta12a,wang13a,sanchez15a}. These systems have been connected to a number of potential applications, including: cancer treatment~\cite{nelson10a,wang14a}, drug delivery~\cite{kagan10a,sundararaj10a,gao12a,gao14a}, soil remediation~\cite{lien99a}, and microfluidic mixing~\cite{kim04a,hernandez05a,kim07a,pushkin13}. In addition, swimmers show promise as model systems for out-of-equilibrium phenomena~\cite{cates12a,cates15a}, for which a thermodynamic description is still in its infancy when compared to the fully established formalism of statistical physics. 

A particular class of active colloids is often connected with these potential applications, namely, artificial self-phoretic colloids~\cite{brown14a,ebbens12a,ebbens14a,howse07a,lee14a,paxton04a,simmchen14a,valadares10a,wang06a}. This is because their fabrication can be well controlled and they appear to be simpler than biological organisms. Self-phoretic colloids are propelled by means of self-generated fields of solute molecules that interact with the colloid. These gradients are typically caused by chemical decomposition reactions that take place on the surface of the particle. The most common systems that exploit self-phoresis are Au-Pt nanoparticles~\cite{lee14a,paxton04a,wang06a} and Pt-coated Janus spheres~\cite{brown14a,ebbens12a,ebbens14a,howse07a,simmchen14a,valadares10a} that decompose hydrogen peroxide into water and oxygen. The former are considered self-electrophoretic~\cite{moran10a,paxton04a,sabass12b}, while for the latter the nature of the phoretic mechanism is still hotly debated~\cite{brown14a,brown15a-pre,ebbens14a}.

It was recently shown theoretically by Brown~\textit{et al.}~\cite{brown15a-pre} that bulk association-dissociation reactions of the solutes involved in the surface reactions -- the surface reactions drive the system out-of-equilibrium -- strongly impact the speed of these particles. Thus, it is necessary to always consider a possible electrophoretic component in the self-propulsion of all current experimental (aqueous) systems. Moreover, Ref.~\cite{brown15a-pre} underpins the poor understanding of the surface reactions that take place in experimental systems, which makes the current modelling of swimmers rather tentative. 

The main issue is that the local surface fluxes of the various species (including reaction intermediates) cannot be straightforwardly measured. Only the total surface reaction rate has been established~\cite{brown14a}. However, it has been shown that there is a dependence of the reactivity on the thickness of the Pt-coating~\cite{ebbens14a}. The vapour-deposition procedure, by which most Janus particles are created, causes a nonuniform thickness of platinum from the equator to the pole of the Janus swimmer~\cite{ebbens14a}. Therefore, it is likely that there is a heterogeneous reactivity on the particle's surface. A systematic study of the effect of such a reaction heterogeneity has not yet been performed. However, the recent investigation into the effect of shadowing a Au-coated Janus swimmer with a Pt-patch~\cite{gibbs10a} provides insights in how to carry out such a study experimentally. 

In this paper, we theoretically consider the effect of surface-reaction heterogeneity on the properties of a self-phoretic swimmer. We consider the `swimming efficiency', which is the hydrodynamic power output over the total enthalpy produced by the reaction, in order to compare different systems on an equal footing. This definition was first introduced by Paxton~\textit{et al.} in the footnotes to Ref.~\cite{paxton05a}. The concept of the efficiency of swimming has since been investigated by Sabass and Seifert, who showed that nanoparticles are far more efficient at self-propulsion than micron sized colloids~\cite{sabass10a} and examined this quantity the context of self-electrophoresis~\cite{sabass12a}. More recently, Wang~\textit{et al.}~\cite{wang13b} performed an analysis of the efficiency of various types of swimmers.

We use the swimming efficiency to demonstrate that there is a strong dependence of the swimming speed on the specifics of the surface-reaction heterogeneity, given a total overall reaction rate. This is unsurprising in light of the work of Popescu~\textit{et al.}~\cite{popescu10a}, but our swimming efficiency allows us to compare these different distributions on an equal footings. We can show that the most favourable way to propel a particle is by localizing the chemical reactions on tiny (isolated) spots on the surface. That is, on a single spot in the case of self-diffusiophoresis, and on two diametrically opposing (absorbing and emitting) spots (or poles) in the case of self-electrophoresis. This maximizes the prefactor in the first Legendre mode of the solute distribution, that determines the swimming speed. Such polar configurations have been considered before~\cite{golestanian07a,popescu10a}, but have not been connected to efficiency or been systematically compared. Popescu~\textit{et al.}~\cite{popescu10a} demonstrate that the polar distributions result in a zero swimming speed. However, in their work, the total reaction rate is proportional to the catalytic surface area, whereas our swimmers have a finite total reaction rate independent of the catalytic distribution.

Our result is of interest for the fabrication of colloidal swimmers, as it implies that significant swimming speeds may be achieved by using a minimal amount of reactive material. The reduced surface area of a `point-like' particle at the pole does not necessarily imply a strongly reduced chemical decomposition rate, as it was recently hypothesized that the fast swimming of nanoparticles might be explained by a relatively large catalytic rate compared to that of a micron-sized colloid~\cite{brown15a-pre}. This is particularly true in the diffusion-limited regime~\cite{ebbens12a}, where the overall reaction rate is determined by the rate at which fuel molecules can diffuse in from the bulk. Another advantage of the polar distribution is that metal surfaces, such as the Pt coating, lead to strong van der Waals interactions between swimmers that can cause the swimmers to aggregate~\cite{buttinoni13a}. By localizing the metallic reactive site, such aggregation may be suppressed. In addition, for the polar swimmers more of the surface is available for chemical modification that is not related to generating self-propulsion, e.g., for binding with cancer cells or chemically decomposing pollutants. Finally, the flow field around the particle is strongly modified with respect to that of a hemispherically coated object, which could lead to enhanced microfluidic mixing. The experimental realization of such particles, however, remains an open problem. There are indications that such a localized reactivity can be achieved~\cite{valadares10a}, but obtaining localized high reaction rates could prove challenging.

The remainder of this paper is structured as follows. In Sections~\ref{sec:methods_fem} we introduce the self-diffusiophoretic and self-electrophoretic model and discuss their numerical solution using a finite-element method; details of the theoretical analysis may be found in Appendices~\ref{app:model_dp} and~\ref{app:model_el}, respectively. Next, we introduce the swimming efficiency in Section~\ref{sec:efficiency}. This is followed by a presentation of our results in Section~\ref{sec:results}. Finally, we give a conclusion and present an outlook in Section~\ref{sec:conclusion}.

\section{\label{sec:methods_fem}Finite Element Simulations}

In this manuscript we consider both self-diffusiophoretic and self-electrophoretic swimmers. For convenience we restrict ourselves to spherical swimmers of radius $a$ that are axisymmetric in the $z$-axis. Here, we limit ourselves to a minimal description of both types of swimmers in an aqueous environment. Both swimmers decompose hydrogen peroxide catalytically on their surface, which drives the system out-of-equilibrium. This is modelled using a flux boundary condition $f(\theta) = c_J f_J(\theta)$, which only depends on the polar angle $\theta$. The dimensionful constant $c_J > 0$ bears all the units while the dimensionless function $f_J(\theta)$ gives the magnitude of the local production rate. If the production rate is negative, the species is instead consumed by the surface reaction. The motion of the swimmer can be determined by solving the coupled system of linear differential equations consisting of: the Stokes, Nernst-Planck, and Poisson equation. Full details of the two models are provided in Appendices~\ref{app:model_dp} and~\ref{app:model_el}, respectively. 

For the purposes of computing swimming efficiency the following information is required: (i) Self-diffusiophoresis can be described using a single solute species, namely the oxygen,~\cite{popescu10a} which has zero bulk concentration. The swimmer moves due to short-ranged (non-electrostatic) interactions between the oxygen and the swimmer surface, which may be captured within the slip-layer approximation~\cite{anderson89a, golestanian07a, popescu10a, brady11a, sharifimood13a, ebbens14a, shklyaev14a, degraaf15a, michelin15a}. Since oxygen is nowhere consumed on the particle we can choose $f_J(\theta)\geq 0$. (ii) For the self-electrophoretic swimmers, we consider the standard Au-Pt model of Refs.~\cite{paxton04a,moran10a,sabass12b,brown15a-pre}, in which reactions at the surface produce and reduce an ionic species. This sets up a current of charged solutes and an electric field which induces motion of the swimmer which has a surface charge $\sigma$. A minimal model considers only a flux of hydrogen ions H$^{+}$, which are created on one end of the swimmer and reduced on the other, and must have a finite bulk concentration $c_{\mathrm{a}}^{\infty}$ (the subscript `a' stands for active) to produce a finite swimming speed. Simultaneously, a current of electrons flows through the swimmer to ensure charge conservation. Since H$^{+}$ is both formed and reduced, there is no restriction on the sign of $f_J(\theta)$. However, charge conservation requires $\int_{-1}^1 \tilde{f}_J(b) \mathrm{d}b = 0$, where we have introduced $b=\cos\theta$ the unit arc length and $\tilde{f}_{J}(b) \equiv f_J(\theta)$. In the self-electrophoretic model, there are also two ionic species that induce electrostatic screening with Debye length $\kappa^{-1}$, with bulk concentrations $c_{\pm}^{\infty}$, but which are not involved in the reaction; so-called spectator species. 

\begin{figure}[]
\centering
\includegraphics[width=8.5 cm]{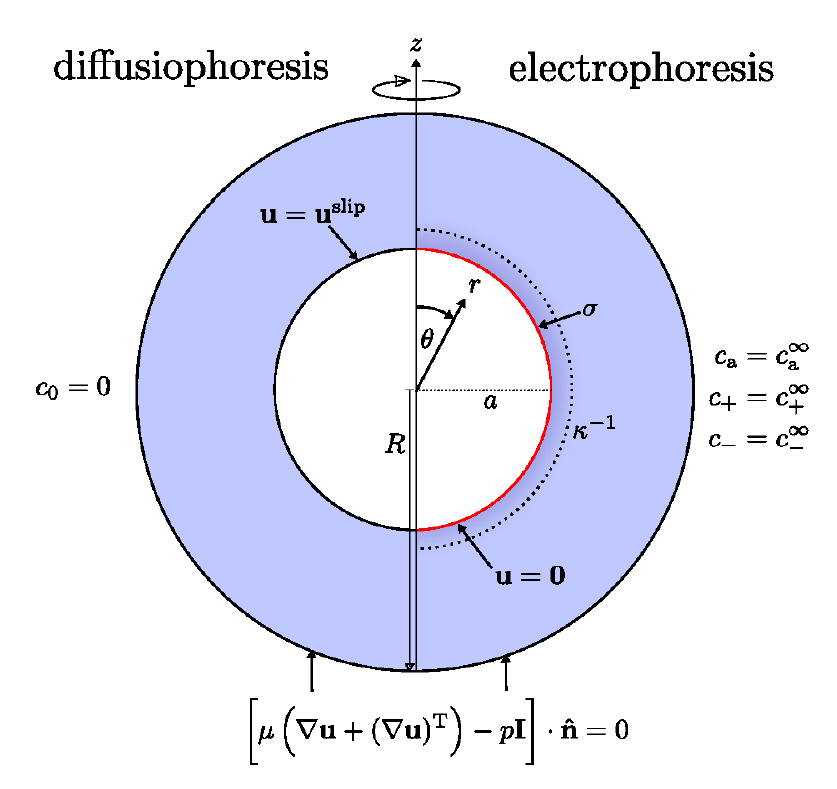}
\caption{\label{fig:schematic_fem} Schematic of the finite element simulation setup for a swimmer of radius~$a$. (Left) the diffusiophoretic case, (right) the electrophoretic case. The setups are rotationally symmetric around the $z$-axis. On the edge of the simulation domain of radius~$R$ the concentrations of all species are set to the bulk concentration and the fluid flow field has to fulfill the no-stress boundary condition. For diffusiophoresis a slip-velocity used on the swimmer's surface. For electrophoresis a no-slip boundary condition is applied. The surface charge density~$\sigma$ leads to formation of a screening layer of Debye length~$\kappa^{-1}$.}
\end{figure}

Appendices~\ref{app:model_dp} and~\ref{app:model_el} derive the analytic expressions for the velocity of the two types of swimmer in the linearized regime (for small solute flux and low surface charge). In addition to our analytic work, we solve the full (nonlinear) equations using the finite element method (FEM); here we limit ourselves to the linear regime to verify our calculations. Using the COMSOL Multiphysics Modeling software, we obtain numerical solutions to the respective system of equations via the procedures outlined in \cite{brown15a-pre,degraaf15a}. A schematic of the simulation setup is shown in Fig.~\ref{fig:schematic_fem}. The rotational symmetry of the problem makes it possible to work on a quasi-two-dimensional axisymmetric domain. In our numerical calculations we also neglect the advective coupling Nernst-Planck equation, which allows us to treat the diffusion/electrostatics part and the hydrodynamics part of the overall problem separately~\cite{brown15a-pre}. This is reasonable since for relevant experimental systems the P{\'e}clet number, a dimensionless quantity that is defined as the ratio of the rate of advection and the rate of diffusion for transport processes in hydrodynamic systems, $\mathrm{Pe} \le 10^{-2}\ll 1$. A major difference to the theoretical calculation is that for the flow field we apply a no-stress boundary condition at the edge of the simulation domain:
\begin{equation}
	\left[ \mu \left( \nabla \vec{u} + (\nabla \vec{u})^\mathrm{T} \right) - p \vec{I} \right] \cdot \hat{\vec{n}} = 0, \label{equ:no-stress}
\end{equation}
where $\vec{u}$ is the fluid flow velocity, $p$ the pressure, $\mu$ the dynamic viscosity, and $\hat{\vec{n}}$ the unit normal to the boundary. In addition, $\cdot$ indicates the inner product, $\nabla$ the gradient, ${}^\mathrm{T}$ transposition, and $\vec{I}$ the identity matrix. This choice allows us to directly determine the swimmer velocity by averaging the fluid velocity over the edge of our simulation domain, rather than solving the problem for this quantity.

Unless otherwise specified, we use the following parameters in our numerical verification. We simulate a (colloidal) swimmer of radius $a = \SI{0.5}{\micro\meter}$ in a simulation domain of radius $R = \SI{5.5}{\micro\meter}$, which is larger than $10 a + 25 \kappa^{-1} \approx \SI{5.2}{\micro\meter}$; the size of the domain found to be adequate in Ref.~\cite{brown15a-pre}. This ensures that the electrostatic potential at the boundary of the simulation domain has decayed sufficiently. We use a temperature of $T = \SI{298.15}{\kelvin}$ (room temperature). The fluid (water) surrounding the swimmer has density $\rho_\mathrm{f} = \SI{1.0e3}{\kilogram\per\cubic\meter}$, viscosity $\mu = \SI{1.0e-3}{\kilo\gram\per\meter\per\second}$, and relative permittivity $\epsilon_r = 78.36$.

In the diffusiophoretic model, the net flux of solute molecules through the surface of the swimmer is ${\cal F} = \SI{5.0e-13}{\mole\per\second}$, to ensure that we are in the linear regime. The diffusivity of the active solute is $D_0 = \SI{1.9e-9}{\square\meter\per\second}$ for oxygen~\cite{han96a,sridhar83a}. The interaction between solute molecules and the swimmer surface is given by the slip-layer parameter $\xi = \SI{-1.0d-15}{\meter\tothe{5}\per\mole\per\second}$, see Appendix~\ref{app:model_dp}. These last two parameters are choices that enable us to compute the speed in the simulation and are based on the values in Ref.~\cite{degraaf15a}, but otherwise do not affect our results for the swimming efficiency. 

For the self-electrophoretic swimmers we use parameters from Ref.~\cite{brown15a-pre} in order to model a Au-Pt swimmer in H$_{2}$O$_{2}$ with added NaCl. The active charge carrying species is H$^{+}$ in this case. The concentrations of spectator ions are $C_{+} = \SI{1.0e-3}{\mole\per\liter}$ and  $C_{-} = \SI{1.001e-3}{\mole\per\liter}$. The bulk concentration of the active species is $C_\mathrm{a} = \SI{1.0}{\micro\mole\per\liter}$, which has to be nonzero to ensure a finite swimming velocity. The diffusivities of the three species are $D_\mathrm{a} = \SI{9.3e-9}{\square\meter\per\second}$ (H$^{+}$~\cite{haynes13a}), $D_+ = \SI{1.3e-9}{\square\meter\per\second}$ (Na$^{+}$), and $D_- = \SI{2.0e-9}{\square\meter\per\second}$ (Cl$^{-}$). The net flux of active species through the swimmers' surface is ${{\cal F} = \SI{1.5e-18}{\mole\per\second}}$, to ensure that the system is in the linear regime. The swimmer has a surface charge $\sigma = \SI{1.0e-4}{\elementarycharge\per\square\nano\meter}$, with \si{\elementarycharge} the elementary charge. For these choices the Deybe length has a value given by $\kappa^{-1} = \SI{9.6}{\nano\meter}$.

\section{\label{sec:efficiency}Swimmer Efficiency}

To have a dimensionless measure for the efficiency of spherical swimmers with different heterogeneous surface reactivities and radii~$a$, we define the swimmer efficiency functional
\begin{equation}
	\eta = \frac{6\pi \mu a U^2}{{\cal F}^\mathrm{diff/el} \epsilon_\mathrm{ch}}, \label{equ:def_efficiency}
\end{equation}
with $U$ the propulsion speed and $\epsilon_\mathrm{ch}$ the Gibbs-free-energy change on production of a single solute molecule. Here, we have followed Refs.~\cite{paxton05a,sabass12a,wang13b} by considering an equivalent particle that is dragged by an external force through the fluid at speed $U$ to compute the energy dissipation by the active particle (power output). The total reaction rate $\cal F$ in Eq.~\eqref{equ:def_efficiency} is given by
\begin{align}
  \cal F^{\mathrm{diff}} &\equiv 2 \pi a^{2} \int_{-1}^1 \tilde{f}_{J}(b) \mathrm{d} b \label{eq:diff_totflux}; \\
  \cal F^{\mathrm{el}} &\equiv \pi a^{2} \int_{-1}^1 \vert \tilde{f}_{J}(b) \vert \mathrm{d} b \label{eq:el_totflux},
\end{align}
For the electrophoretic case, a factor of a half is introduced, since for each outgoing (production; $\tilde{f}_{J}(b)>0$) and incoming (consumption; $\tilde{f}_{J}(b)<0$) active solute, only one electron travels through the swimmer. That is, both steps are required to complete the reaction and liberate only one $\epsilon_\mathrm{ch}$ together. 

We now insert the swimming speeds that we theoretically obtained (Eqs.~\ref{equ:diff_propulsion_speed} and~\ref{equ:el_propulsion_speed}, respectively) into the efficiency functional. This allows us to write
\begin{align}
	c_\eta^\mathrm{diff} &\equiv \frac{3}{4} \frac{\mu c_J \xi^2}{a D_0^2 \epsilon_\mathrm{ch}},\\
	c_\eta^\mathrm{el} &\equiv \frac{3}{2} \frac{c_J}{a \mu \epsilon_\mathrm{ch}} \left( \frac{\sigma e z_\mathrm{a}}{D_\mathrm{a} \kappa ^3 \varepsilon} H(\kappa a) \right)^2, \label{equ:c_eta_el}
\end{align}
where $z_\mathrm{a}$ is the valency of the active species and $H(\kappa a)$ is the generalized Henry function~\cite{brown15a-pre} (for the FEM parameters $H(\kappa a) = \SI{0.83}{}$), also see Appendix~\ref{app:model_el}. We can then express the swimmer efficiency as
\begin{equation}
	\eta = c_\eta^\mathrm{diff/el} \frac{\left(  \int_{-1}^1 b \tilde{f}_J(b) \mathrm{d} b \right)^2}{\int_{-1}^1 |\tilde{f}_J(b)| \mathrm{d} b}, \label{equ:comb_eta}
\end{equation}
where it should be remembered that $\tilde{f}_J$ has different constraints for the two mechanisms of self-phoresis that we consider.

To determine the maximum possible swimmer efficiency we assume $\tilde{f}_J(b)$ to be normalized, that is
\begin{equation}
{\int_{-1}^1 |\tilde{f}_J(b)| \mathrm{d} b} = 1.
\end{equation}
This can be done without loss of generality, since the definition of the production function (Eq.~\ref{equ:production_f}) allows any normalization constant to be absorbed into the choice of $c_J$. Using
\begin{align}
	\left| \int_{-1}^1 b \tilde{f}_J(b) \mathrm{d} b \right| \leq  \int_{-1}^1\left| b \right| \left|\tilde{f}_J(b)\right| \mathrm{d} b \leq \int_{-1}^1 \left|\tilde{f}_J(b) \right| \mathrm{d} b,
\end{align}
it is easily shown, that an upper limit on the swimmer efficiency is given by
\begin{align}
	\eta_\mathrm{max} &= c_\eta^\mathrm{diff/el}.
\end{align}
In addition, since the hydrodynamic dissipation cannot exceed the power input due to the reaction, we have that $c_\eta^\mathrm{diff/el}< 1$ always, which ensures the efficiency to be well-defined.

\section{\label{sec:results}Results}

Interpreting the factor $b$ in the integral of the numerator in Eq.~\ref{equ:comb_eta} as a weighting factor for the production function~$\tilde{f}_J$, suggests that maximum swimmer efficiency can only be obtained for delta-distribution-like production functions at the pole(s) of the swimmer. That is, the production function's value is reduced by the factor $\vert b \vert <1$ for all polar angles, except for $\theta \in \{0,\pi \}$. It is therefore impossible to reach maximum efficiency if there are contributions to the flux away from the poles.

To verify this for the self-diffusiophoretic case, we assume a production function
\begin{equation}
	\tilde{f}_J(b) = \pm\delta(b-\alpha),
\end{equation}
where \mbox{$\alpha \in [-1,1]$}. Then, the corresponding efficiency functional yields
\begin{equation}
	\eta =\left|\pm \alpha c_\eta^\mathrm{diff} \right| \leq \left|c_\eta^\mathrm{diff} \right| = \eta_\mathrm{max}.
\end{equation}
As expected, maximum swimmer efficiency is obtained, if all the solute production or consumption happens at an isolated point on the swimmer surface. For the self-electrophoretic case, we impose the production function
\begin{equation}
	\tilde{f}_J(b) = \delta(b-\alpha)-\delta(b+\alpha),
\end{equation}
which leads to the efficiency functional
\begin{equation}
	\eta =\left|\alpha c^\mathrm{el}_\eta \right| \leq \left|c^\mathrm{el}_\eta \right| = \eta_\mathrm{max}.
\end{equation}
Maximum swimmer efficiency is obtained if all the solute production occurs at an isolated point on the swimmer surface, while all the consumption happens at an isolated point on the opposite side of the colloid. Here it is worth noting that Popescu~\textit{et al.}~\cite{popescu10a} found the polar distribution for self-diffusiophoresis to give zero swimming speed for self-diffusiophoresis. However, their total reaction rate is proportional to the surface area, whereas ours is constant.  

\begin{figure*}[]
\centering
\includegraphics[width=17.5 cm]{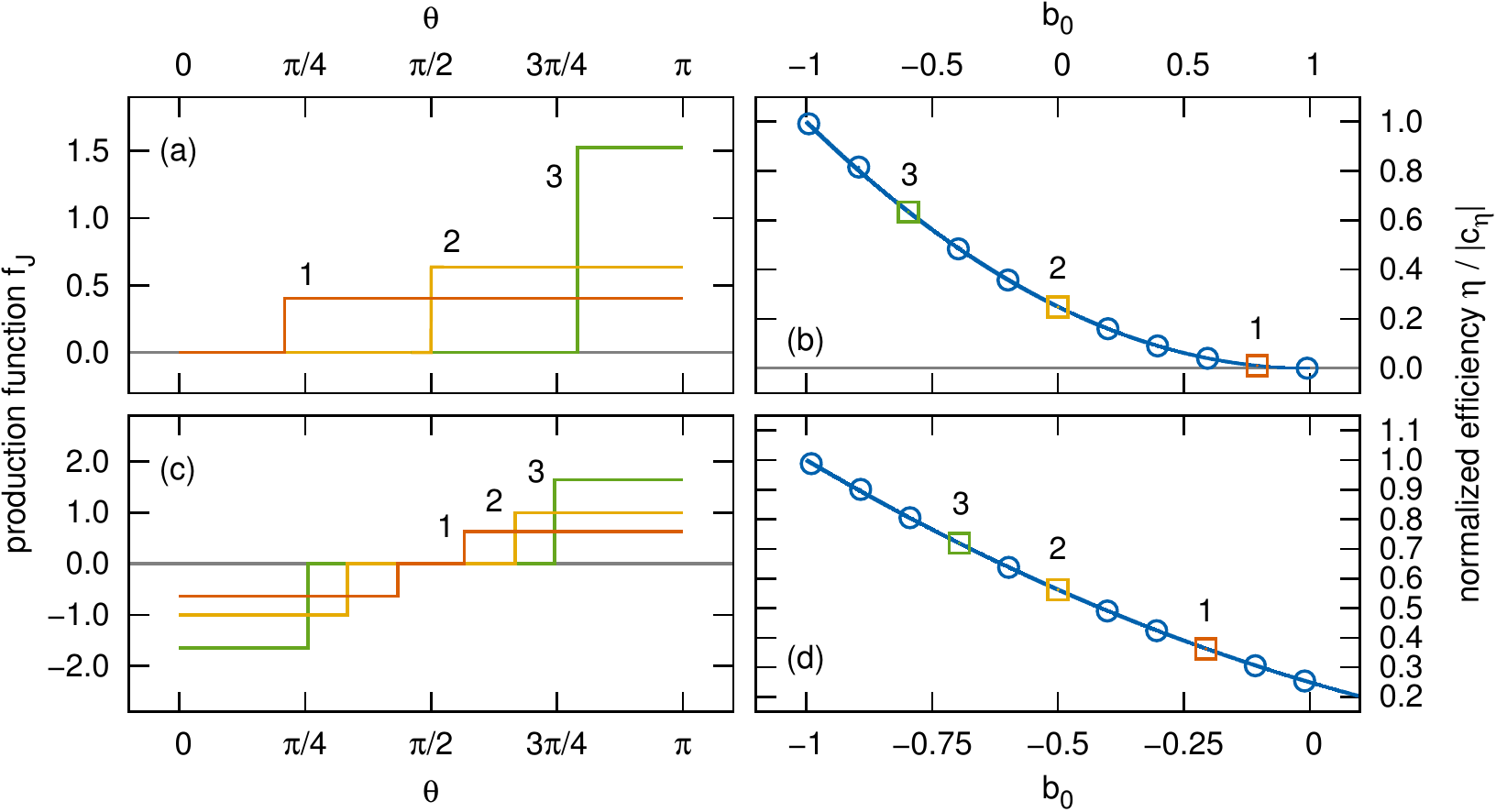}
\caption{\label{fig:diffel_plot} The swimming efficiency for a step function-type flux profile in the diffusiophoretic (a,b) and the electrophoretic model (c,d). The left-hand side shows example surface flux profiles $f_J$ as a function of the polar angle $\theta$. The right-hand side gives the normalized efficiency \mbox{$\eta / |c_\eta|$} as a function of $b_0$ obtained from theory (blue curve) and FEM simulations (circles). The coloured squares indicate the points that correspond to the profiles shown on the left.}
\end{figure*}

To support our findings, we performed both analytical and numerical calculations for different classes of surface production profiles, of which we show a sample here in Fig.~\ref{fig:diff_plot}; we refer to Appendix~\ref{app:sensi} for additional curves. Figure~\ref{fig:diff_plot} shows various step-function-type $f_{J}$ and the resulting efficiency for both the self-diffusiophoretic and self-electrophoretic model. The step functions are characterized by a cut-off angle $\theta_0=\acos(b_0)$. Up to this angle we assume the production function to be $f_J(\theta<\theta_0)=0$. For larger angles the production function has a constant value. The efficiency functionals for the electrophoretic model were chosen to be antisymmetric; however, non-antisymmetric profiles are in principle allowed, provided the net flux is zero.  

We find excellent agreement between the analytic and FEM solutions, indicating that our results are trustworthy. For both, the self-diffusio- and the self-electrophoretic model, we observe the expected behaviour, that is, the normalized efficiency functional converges to $1$ when the production functions approach the delta-like form described above. Our results also show the sensitive dependence of a swimmer's speed on its production function. The typical half-coated diffusiophoretic Janus swimmer ($\theta_{0} = \pi/2$, profile \#2 in Fig.~\ref{fig:diffel_plot}a) has an efficiency of $\eta = 0.25 \eta_{\max}$ and achieves $\vert U/U_{\max} \vert = 0.50$, with $U_{\max}$ the speed of the polar swimmer, for the parameters of Section~\ref{sec:methods_fem}. A ramp or inverse-ramp swimmer ($\theta_{0} = \pi/2$) achieves $\vert U/U_{\max} \vert = 0.69$ ($\eta=0.47 \eta_{\max}$) and $0.39$ ($\eta=0.15 \eta_{\max}$), respectively, see Appendix~\ref{app:sensi}. The swimming efficiency of the ramp case is clearly the largest, but the speed is still less than half of the theoretical maximum. This simple example shows that the speed of swimming (and efficiency) can be substantially improved by modifying (localizing) the reactivity, under the condition of constant $\cal F^{\mathrm{diff/el}}$.

\begin{figure}[]
\centering
\includegraphics[width=8.5 cm]{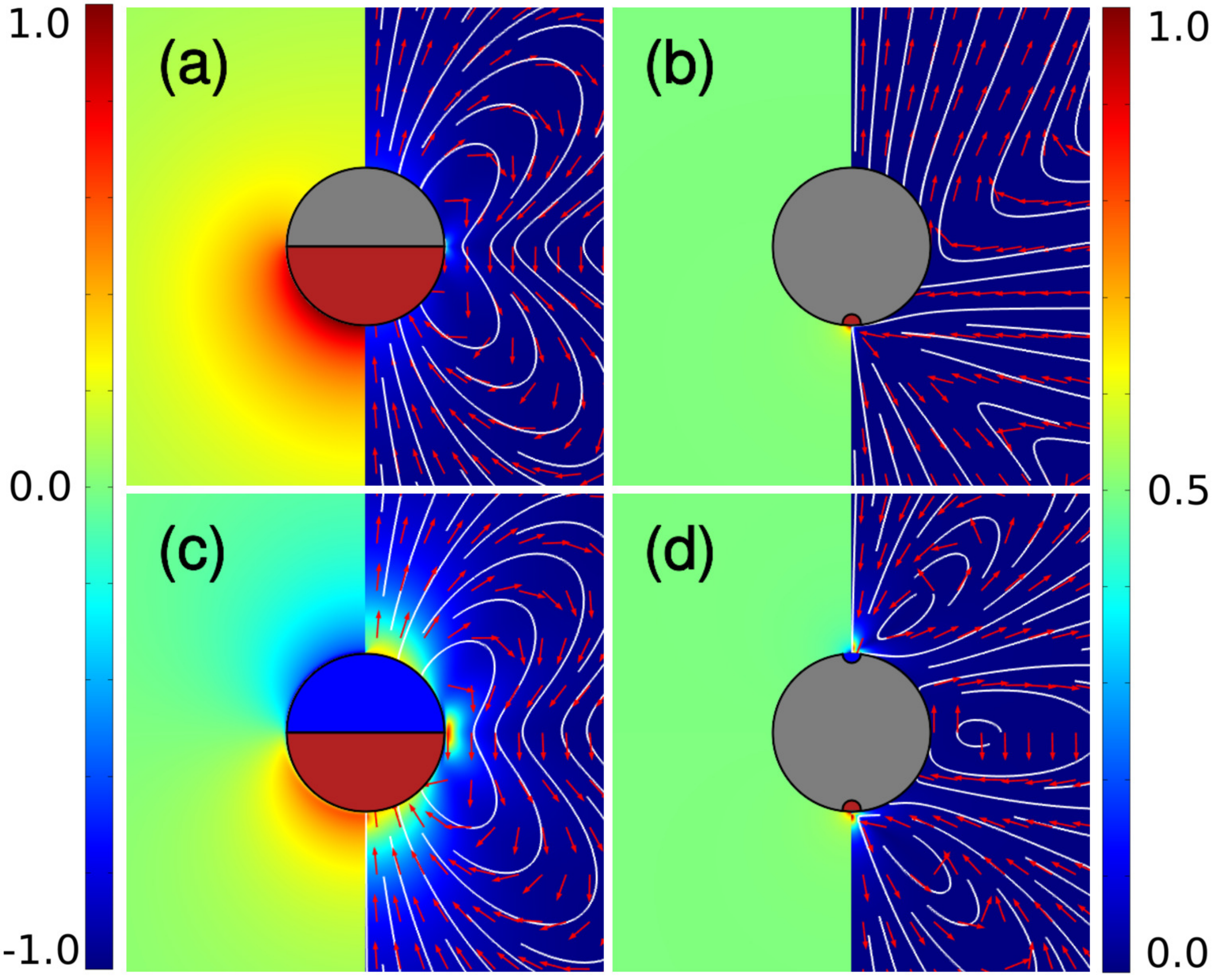}
\caption{\label{fig:fields_fem}Normalized concentration field (left) and normalized fluid flow field (right) in the lab frame for self-diffusiophoretic swimmers (a,b) and self-electrophoretic swimmers (c,d). Figures (a,c) show simulation results for half-coated swimmers with a Heaviside flux ($\theta_{0} = \pi/2$) production profile and (b,d) show swimmers with delta distribution like production profiles. The white lines are stream lines, the normalized red arrows depict the flow field, but not its amplitude.}
\end{figure}

Next, we consider the differences in the concentration field and the fluid flow field between the delta-type and half-coated (Heaviside) swimmer, see Figure~\ref{fig:fields_fem}. The first thing to note is that the flow field for the delta-type self-diffusiophoretic colloid has the shape of a (slightly-perturbed) Stokes dipole, whereas the Heaviside swimmer has a source dipole flow field with a strong perturbation near the equator. This is because the Heaviside swimmer is completely antisymmetric with respect to the equator, so it does not have pusher/puller components in the flow field, whereas the delta-type swimmer is not antisymmetric. The flow field of the delta-type self-electrophoretic swimmer is strongly quadrupolar in nature. 

The way the flow field around the particle is modified for the delta-type flux profiles can have advantages. In particular, the interaction between particles is strongly influenced by these flow fields. The fact that the diffusiophoretic swimmer (puller, in this case) has only a tiny metallic spot and is pushing fluid away from this spot, makes it difficult for particles to irreversibly aggregate due to van der Waals forces between the metal surfaces. Analysis of the decay of the flow field shows that the delta-type swimmers have much longer ranged hydrodynamic flow fields. This can be easily understood, as the Heaviside swimmer's flow field has larger contributions of the higher-order modes due to the asymmetry, which exhibit stronger decay. Therefore, the energy that is available is dissipated over a smaller range. This can significantly impact the ability of the swimmer to mix its environment. In addition, the strongly quadrupolar flow field of the self-electrophoretic delta-type swimmer could also lead to enhanced stirring of the fluid~\cite{dunkel10a}.

\section{\label{sec:conclusion}Discussion and Outlook}

Summarizing, we have considered the influence of the surface reactivity of an active colloid on the speed and efficiency of phoretic self-propulsion. In order to do so, we used the so-called swimmer efficiency~\cite{paxton05a,sabass12a,wang13b}, which is the swimmer's hydrodyamic power output over the total enthalpy produced by the reaction. We find that the most efficient way to self-propel is to have the surface reactions take place in an isolated spot on the surface of the pole in self-diffusiophoresis. In self-electrophoresis, the most efficient form is achieved by two diametrically opposite spots (poles) that put out and absorb charged species, respectively. This is verified theoretically and numerically (using finite elements) by considering various reactivity profiles. The study further underscores the sensitive nature of the speed and flow profiles on the reactivity distribution and the limited insight that can be gained from the total surface flux and speed alone. 

The polar distributions of reactivity have several advantages over the typical hemispherical surface distributions; however, it requires potentially impossible reaction rates per unit area. First, this distribution makes optimal use of the chemical reactions, while at the same time requiring a minimal amount of reactive material.  Second, more of the surface is available for functionalization that is not related to achieving self-propulsion. Third, the limited surface that is covered by metallic catalysts, would facilitate suppression of microswimmer aggregation. Finally, the flow fields for polar distributions are strongly dipolar and octapolar for the self-diffusiophoretic and self-electrophoretic swimmers, respectively. This may be beneficial for their interaction with surfaces and their ability to stir the surrounding fluid. 

Future work will focus on the important question of properly defining the efficiency of active colloids in a wider, system aspecific context. The efficiency of the current artificial self-propulsion mechanisms is widely considered to be low for micron-size particles~\cite{paxton05a,sabass10a,sabass12a,wang13b}. Understanding where the conversion bottleneck lies and how biological swimmers manage to be more efficient is instrumental in achieving real world applications. However, this necessitates a formalism that allows for comparison between the vastly different mechanisms utilized in biology and man-made applications, the specific form of which is unclear at this point. The experimental realization of our suggested polar-reactive particles is another open problem, as achieving high reaction rates on a small site is likely experimentally challenging. Nevertheless, the possibility of more efficient use of fuel, as well as a host of other potential benefits, make polar-driven colloids an avenue worth pursuing.

\section*{\label{sec:acknowledgements}Acknowledgements}

JdG gratefully acknowledges  financial  support  by  an NWO Rubicon Grant (\#680501210) and funding by a Marie Sk{\l}odowska-Curie Intra European Fellowship (G.A. No. 654916) within Horizon 2020. We thank the DFG for funding  through  the  SPP 1726  ``Microswimmers  --  From Single Particle Motion to Collective Behaviour''. We would also like to thank A. Brown  for  useful  input  concerning the modeling of the self-electrophoretic swimmers.

\appendix

\section{\label{app:model_dp}The Self-Diffusiophoretic Model}

To study self-diffusiophoresis, we consider a spherical microswimmer of radius~$a$ surrounded by a fluid which contains only one solute indicated using the subscript 0; we will come back to this choice shortly. We use the standard model of self-diffusiophoresis, which we will briefly summarize here. Given the concentration field of the solute $c_0$ and its diffusivity $D_0$, the solute flux $\vec{j}_0$ is
\begin{equation}
	\vec{j}_0 = \vec{u} c_0 - D_0 \nabla c_0 - \frac{D_0 c_0}{k_\mathrm{B} T} \nabla\Psi^\mathrm{diff}, \label{equ:diff_flux_lab}
\end{equation}
with $\vec{u}$ the fluid velocity, $k_\mathrm{B}$ Boltzmann's constant, $T$ the temperature, and $\Psi^\mathrm{diff}$ the interaction potential between solute molecules and the swimmer's surface. The fluid flow field satisfies the Stokes equation, as well as the incompressibility condition
\begin{align}
	\mu \nabla^2 \vec{u} &= \nabla p + \vec{f}, \label{equ:stokes} \\
	\nabla \cdot \vec{u} &= 0, \label{equ:incomp}
\end{align}
with fluid dynamic viscosity $\mu$, hydrostatic pressure $p$, and force density $\vec{f} = c_{0} \nabla \Psi^\mathrm{diff}$. We only consider the time-independent case, in the low-Reynolds number regime. The dimensionless Reynolds number is defined as $\mathrm{Re} = \rho_\mathrm{f} v L / \mu$, with $\rho_\mathrm{f}$ the mass density of the fluid, $v$ the maximum velocity of the swimmer relative to the fluid, and $L$ the size of the swimmer. A low Reynolds number indicates that inertial forces in the system are small compared to viscous forces. Also of relevance is the P{\'e}clet number, a dimensionless quantity that is defined as the ratio of the rate of advection and the rate of diffusion for transport processes in hydrodynamic systems, that is $\mathrm{Pe} = u L/D$ with $u$ the velocity given by the fluid flow field and $D$ the diffusivity. Assuming a low P{\'e}clet number, typically $\mathrm{Pe} \le 10^{-2} \ll 1$ for swimmers, the advective term in Eq.~\ref{equ:diff_flux_lab} can be neglected, giving
\begin{equation}
	\vec{j}_0 = - D_0 \nabla c_0 - \frac{D_0 c_0}{k_\mathrm{B} T} \nabla\Psi^\mathrm{diff}. \label{equ:diff_flux_co-moving}
\end{equation}

We model the catalytic decomposition of hydrogen peroxide on the swimmer's surface, as a production of a single species of solute molecules. This is permitted for low Dahmk{\"o}hler numbers ($\mathrm{Da} = R L/ D_{0}$, with $R$ the reaction rate)~\cite{popescu10a}. The production rate, which corresponds to the distribution of the reactivity of the catalyst on an actual microswimmer, is assumed to be axisymmetric in the $z$-axis. This is a reasonable reduction for particles created using vapour deposition~\cite{baraban12a,kreuter13a}. Thus, in spherical polar coordinates the production rate can be described by a function
\begin{equation}
	f(\theta) = c_J f_J(\theta), \label{equ:production_f}
\end{equation} 
which only depends on the polar angle $\theta$. Here, the dimensionful constant $c_J > 0$ bears all the units while the dimensionless function $f_J(\theta)$ gives the magnitude of the local production rate. The single species that we model is nowhere consumed on the particle, therefore we can choose $f_J(\theta)\geq 0$. Thus, we have the following boundary condition for the normal solute flux through the swimmer's surface
\begin{equation}
	\vec{j}_0 \cdot \hat{\vec{n}} |_s = c_J f_J(\theta),
\end{equation}
where $\hat{\vec{n}}$ is the unit normal out of the particle surface and $|_s$ indicates evaluation at the surface. In the bulk, no reactions take place, that is
\begin{equation}
	\nabla \cdot \vec{j}_0 = 0.
\end{equation}

The fluid is assumed to have infinite extent with uniform solute concentration $c^{\infty}_0 = 0$ far from the swimmer. There, we also set \mbox{$p \rightarrow p^\infty$}, the atmospheric pressure; the fluid velocity approaches \mbox{$ \vec{u} \rightarrow - \vec{U}$} with $\vec{U}$ the velocity of the swimmer. That is to say, we consider the problem in the frame co-moving with the swimmer, for which $\vec{u}|_{s} = \vec{0}$.

Theoretical analysis shows that when the interaction potential $\Psi^\mathrm{diff}$ decays to zero on a length scale \mbox{$\delta \ll a$}, the force terms in the fluid velocity can be replaced by an effective slip boundary condition, which replaces the condition $\mathbf{u}|_{s} = \mathbf{0}$~\cite{anderson89a, brady11a, degraaf15a, ebbens14a, golestanian07a, michelin15a, popescu10a, shklyaev14a, sharifimood13a}. The surface slip is given by
\begin{align}
	\vec{u}(\vec{s})\cdot\hat{\vec{n}} &= 0, \\
	\vec{u}(\vec{s})\cdot\hat{\vec{t}} &= - \xi(\vec{s}) \hat{\vec{t}} \cdot \nabla c_0(\vec{s}),
\end{align}
with $\hat{\vec{t}}$ representing the tangent vector to the surface (there are two, but only the $\hat{\vec{\theta}}$ component contributes due to axisymmetry) and
\begin{equation}
	\xi(\vec{s}) = \frac{1}{\mu} \int_0^\infty t \left[ \exp\left( -\frac{1}{k_\mathrm{B} T} \Psi^\mathrm{diff}(\vec{s}+t \hat{\vec{n}}(\vec{s}))\right) - 1 \right] \mathrm{d}t, \label{equ:diff_xi}
\end{equation}
a parameter that takes into account the surface--molecule interaction. The slip-velocity convention is used throughout this manuscript for the self-diffusiophoretic model.

Here, we assume $\xi$ constant over the surface. We use $\xi$ as a parameter rather than concern ourselves with its relation to the interaction potential. However, the exact nature of $\xi$ will turn out not to be relevant for the swimming efficiency as long as it is homogeneous over the surface. This homogeneity allows one to consider only the first Legendre mode of the solute flux to obtain the swimming speed. The propulsion speed of a self-diffusiophoretic particle is given by \cite{degraaf15a}
\begin{align}
	U^\mathrm{diff} &= \frac{c_J \xi}{2 D_0} \int_{-1}^1 b \tilde{f}_J(b) \mathrm{d} b, \label{equ:diff_propulsion_speed}
\end{align}
where we used the substitution \mbox{$b=\cos\theta$} and we use the tilde to indicate the substitution \mbox{$\tilde{f}_{J}(b)\equiv f_{J}(\acos(b)$)} to ease notation. The integral in Eq.~\eqref{equ:diff_propulsion_speed} is the projection of the surface flux onto the first Legendre mode (save a factor 3/2).

\section{\label{app:model_el}The Self-Electrophoretic Model}

A self-electrophoretic swimmer is similar to the self-diffusiophoretic swimmer of Appendix~\ref{app:model_dp}. However, the solute species have long-ranged electrostatic interactions, which necessitates the introduction of the Poisson equation. The slip-layer approximation can still be made, but only for high salt concentrations~\cite{ebbens14a,sabass12b}. A solution for low salt concentrations can also be found~\cite{brown15a-pre}. Here, we follow~\cite{brown15a-pre} by assuming the linear regime, however, we ignore bulk reactions, as these are not crucial for our purposes. The speed is still only determined by the first Legendre mode of the solute flux, other properties (such as bulk reactivity) can be absorbed into prefactors for the efficiency. 

We assume a spherical swimmer of radius $a$, suspended in a fluid which contains three different solutes, indicated by the use of an index \mbox{$i\in\{\mathrm{a},+,-\}$}. The index `a' stands for `active', i.e., the species which is produced or consumed, which also carries a charge. We write $c_i$ for the concentration field and $D_i$ for the diffusivity. Each solute bears an electric charge $q_i= e z_i$, with $e$ the fundamental charge and valency $z_i = \pm 1$. The flux $\vec{j}_i$ of species $i$ is now given by
\begin{equation}
	\vec{j}_i = \vec{u} c_i - D_i \nabla c_i - \frac{D_i z_i e c_i}{k_{\rm{B}} T} \nabla \phi^\mathrm{el}, \label{eq:flux_ep}
\end{equation}
with the electrostatic potential $\phi^\mathrm{el}$, which satisfies the Poisson equation
\begin{equation}
	\nabla^2 \phi^\mathrm{el} = - \frac{\rho_\mathrm{e}}{\varepsilon}, \label{eq:poisson}
\end{equation}
with charge density
\begin{equation}
	\rho_\mathrm{e} = e \sum_i z_i c_i,
\end{equation}
and the solvent permittivity~$\varepsilon$, which we assume constant. Equation~\eqref{eq:flux_ep} is closed by 
\begin{equation}
	\nabla \cdot \vec{j}_i = 0.
\end{equation}
This means that solute molecules can be produced or consumed only at the swimmer surface. The fluid flow field satisfies Stokes equation (Eq.~\ref{equ:stokes}) and the incompressibility condition (Eq.~\ref{equ:incomp}).

The linearized version of Eqs.~\eqref{eq:flux_ep}~and~\eqref{eq:poisson} are obtained by introducing $y_{0} = e \phi^\mathrm{el}/(k_{\mathrm{B}}T)$ and $y_{i} = (c_{i} - c^{\infty}_{i})/c^{\infty}_{i}$, with $c^{\infty}_i$ the constant, uniform value of the concentration far from the swimmer. We may then write~\cite{brown15a-pre}
\begin{align}
\nabla^{2} y_{i} &= \left\{ \begin{array}{cc} -\frac{e^{2}}{\varepsilon k_{\mathrm{B}} T} \sum_{k} z_{k} c^{\infty}_{k} y_{k} & i=0, \\[0.5em] - z_{i} \nabla^{2} y_{0} & i\in\{\mathrm{a},+,-\} , \end{array} \right.  
\end{align}
for the linearized electrostatic and concentration equations.

A no-slip condition
\begin{equation}
	\vec{u}|_s = 0
\end{equation}
is applied on the colloid's surface. In the bulk, the charge density is
\begin{equation}
	\rho_\mathrm{e}^\infty = e \sum_i z_i c^{\infty}_i = 0.
\end{equation}
We set $\phi^\mathrm{el} \rightarrow 0$ and again $p \rightarrow p^\infty$, the atmospheric pressure, far away. In the co-moving frame, the velocity of the fluid $\vec{u} \rightarrow -\vec{U}$, where $\vec{U}$ is the swimming speed in the laboratory frame. 

Assuming axisymmetry we express the production rate of species `$\mathrm{a}$' -- our active species -- by a function $\tilde{f}(b) \equiv c_{J}\tilde{f}_{J}(b)$ (Eq.~\ref{equ:production_f}). In the steady state an electric current flows through the microswimmer to ensure charge conservation. This requires not only production, but also consumption of active solute molecules at the swimmer surface. Thus, \mbox{$\tilde{f}_J(b)\in\mathbb{R}$} and we require the net normal flux of active species through the surface to be zero (charge conservation) leading to
\begin{equation}
	\int_{-1}^1 \tilde{f}_J(b) \mathrm{d}b = 0. \label{equ:el_f_neutral}
\end{equation}
The production rates of the positively charged (index~$+$) and negatively charged inert species (index~$-$) are zero. However, the presence of these inactive species leads to electrostatic screening, the strength of which is characterized by the Debye length $\kappa^{-1}$, where the inverse Debye length~$\kappa$ is given by
\begin{equation}
	\kappa = \sqrt{\frac{e^2 \sum_i z_i^2 C_i}{\varepsilon k_\mathrm{B} T}}.
\end{equation}

We have the following boundary condition for the normal fluxes through the particle surface
\begin{align}
	\left. \hat{\vec{n}} \cdot \vec{j}_i \right|_s=
	\begin{cases}
		c_J \tilde{f}_J(b), &i=\mathrm{a} \\
		0, &i \in \{+,-\}
	\end{cases}.
\end{align}
For the electrostatic potential we apply a von Neumann boundary condition
\begin{equation}
	\left. \hat{\vec{n}} \cdot \nabla \phi^\mathrm{el} \right|_s = - \frac{\sigma}{\varepsilon},
\end{equation}
which can be applied to a swimmer of vanishing dielectric constant and surface charge density~$\sigma$. In the linear regime this yields the same swimming speed as applying the following Dirichlet (conducting) boundary condition
\begin{equation}
	\left. \phi^\mathrm{el} \right|_s =  \frac{\sigma a}{\varepsilon \left(1 + \kappa a\right)},
\end{equation}
as demonstrated in Ref.~\cite{brown15a-pre}. The linearized versions of the flux and electrostatic Neumann boundary condition read
\begin{align}
\hat{\vec{n}} \cdot \left. \left( \nabla y_{a} + z_{a} \nabla y_{0} \right) \right|_s &= -\frac{c_J }{D_{a}c^{\infty}_{a} }\tilde{f}_J(b) \\
\hat{\vec{n}} \cdot \left. \left( \nabla y_{\pm} + z_{\pm} \nabla y_{0} \right) \right|_s &= 0 \\
\hat{\vec{n}} \cdot \nabla \left. y_{0} \right|_s &= -\frac{\sigma e}{k_{\mathrm{B}} T \varepsilon}.
\end{align}

Our self-electrophoretic model is solved for the velocity of the swimmer following the approach of Brown~\textit{et~al.}~\cite{brown15a-pre}. We do not assume bulk reactions, therefore only regular self-electrophoresis (of the Au-Pt colloid type) is permitted. The swimming speed is then given by
\begin{align}
	U^\mathrm{el} &= - \frac{\sigma c_J e z_\mathrm{a}}{2\mu D_\mathrm{a} \kappa ^3 \varepsilon} H(\kappa a) \int_{-1}^1 b \tilde{f}_J(b) \mathrm{d} b, \label{equ:el_propulsion_speed}
\end{align}
where
\begin{align}
	H(x) = \frac{x^3}{6(x+1)} e^{x} \int_1^\infty  \frac{(t-1)^2 (2 t + 1)}{t^{5}} (1+x t) e^{-t x} \mathrm{d} t
\end{align}
is the generalized Henry function~\cite{brown15a-pre}. Again the projection of the flux onto the first Legendre mode can be recognized in Eq.~\eqref{equ:el_propulsion_speed}. For the cases that we consider in this manuscript $H(\kappa a) = 0.83$.

\section{\label{app:sensi}Sensitivity to the Production Profile}

\begin{figure*}[]
\centering
\includegraphics[width=17.5 cm]{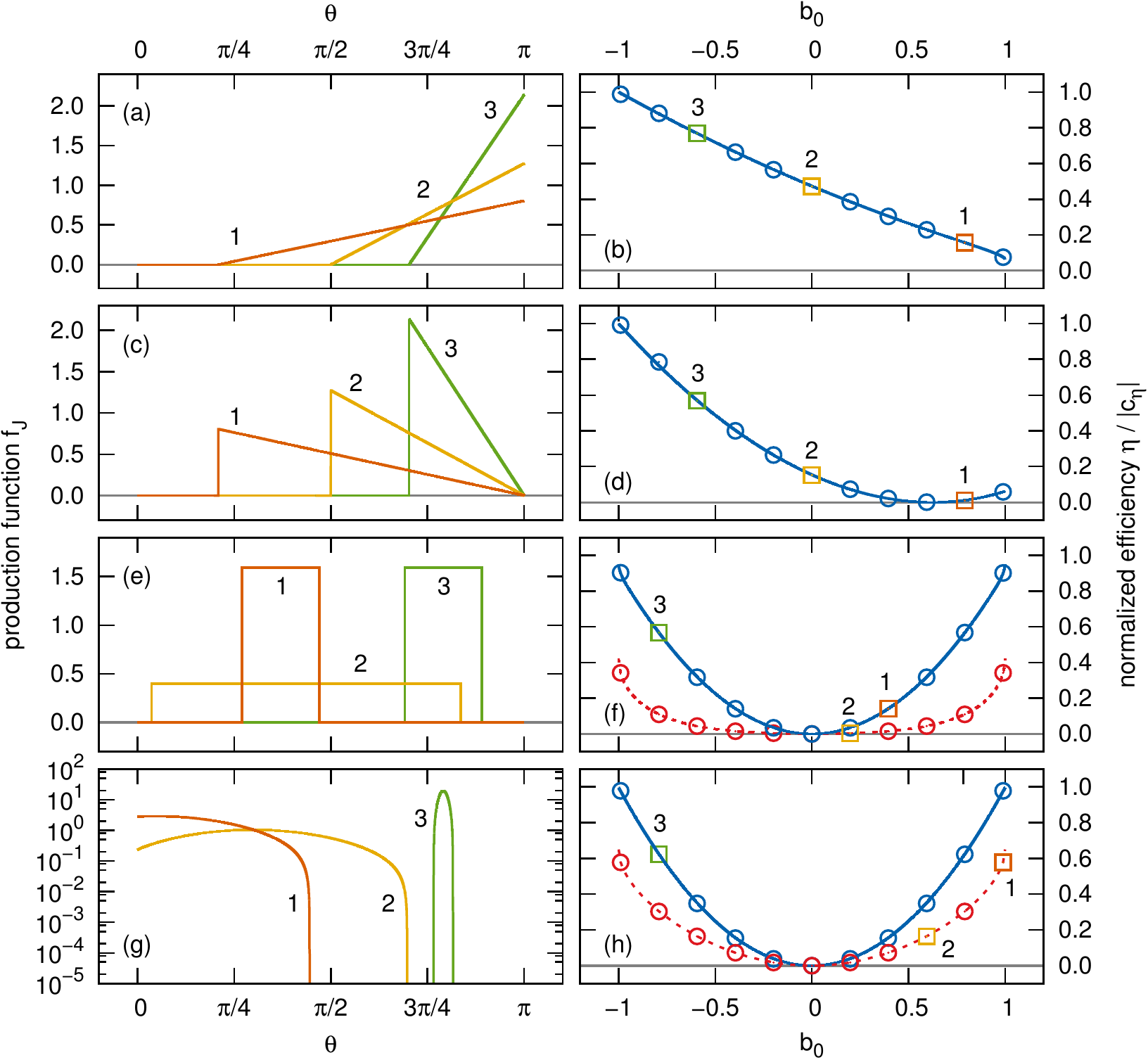}
\caption{\label{fig:diff_plot} The swimming efficiency for a range of surface flux profiles for the self-diffusiophoretic model. The left-hand side shows example surface flux profiles $f_J$ as a function of the polar angle $\theta$: (a) ramps, (c) inverse ramps, (e) a box-like profile, and (g) a shifted Gaussian distribution (Eq.~\ref{equ:shifted_gaussian}; logarithmic scale). The right-hand side gives the normalized efficiency~\mbox{$\eta / |c_\eta|$} as a function of $b_0$ obtained from theory (blue curve) and FEM simulations (circles). The coloured squares indicate the points that correspond to the profiles shown on the left. In (h) and (j) two different widths are considered, $q = 0.05$ (solid blue) and $q = 0.80$ (red dashed).}
\end{figure*}

In addition to the step functions results shown in the main text, we considered ramps and inverse ramps, as well as two slightly more complicated production function profiles, see Fig.~\ref{fig:diff_plot}. The first of which is a box-like shape
\begin{align}
B(\theta) = \frac{1}{q \pi}\Theta(\theta-\theta_0+ q \pi/2) \Theta(-\theta+\theta_0+q \pi/2),
\end{align}
with $\Theta$ the Heaviside step function. The second is a shifted and truncated Gaussian distribution that is smooth. It is non-zero over a width of $4 \sigma$, where $\sigma^2$ is the variance of the distribution:
\begin{align}
G(\theta) =
\begin{cases}\displaystyle
	0, \quad \theta < \theta_0 - \frac{q \pi}{2} \mathrm{~or~} \theta > \theta_0 + \frac{q \pi}{2}\\
	\frac{1}{\sqrt{2\pi}\sigma} \left[ \exp\left( -\frac{(\theta-\theta_0)^2}{2 \sigma^2} \right)-\frac{1}{e^2} \right],~\mathrm{else}
\end{cases} .\label{equ:shifted_gaussian}
\end{align}
These two profile functions are symmetric around $\theta_0$. The width over which the profiles are non-zero is given by $\pi q a$, with $0 \leq q <1$.


\begin{thebibliography}{10}

\bibitem{ramaswamy10a}
S.~Ramaswamy,
\newblock Annu. Rev. Condens. Matter Phys. {\bf 1}, 323 (2010).

\bibitem{marchetti13a}
M.~Marchetti et~al.,
\newblock Rev. Mod. Phys. {\bf 85}, 1143 (2013).

\bibitem{ebbens10a}
S.~J. Ebbens and J.~R. Howse,
\newblock Soft Matter {\bf 6}, 726 (2010).

\bibitem{hong10a}
Y.~Hong, D.~Velegol, N.~Chaturvedi, and A.~Sen,
\newblock Phys. Chem. Chem. Phys. {\bf 12}, 1423 (2010).

\bibitem{sengupta12a}
S.~Sengupta, M.~E. Ibele, and A.~Sen,
\newblock Angew. Chem. Int. Ed. {\bf 51}, 8434 (2012).

\bibitem{wang13a}
W.~Wang, W.~Duan, S.~Ahmed, T.~E. Mallouk, and A.~Sen,
\newblock Nano Today {\bf 8}, 531 (2013).

\bibitem{sanchez15a}
S.~S{\'a}nchez, L.~Soler, and J.~Katuri,
\newblock Angewandte Chemie International Edition {\bf 54}, 1414 (2015).

\bibitem{nelson10a}
B.~J. Nelson, I.~K. Kaliakatsos, and J.~J. Abbott,
\newblock Annu. Rev. Biomed. Eng. {\bf 12}, 55 (2010).

\bibitem{wang14a}
W.~Wang et~al.,
\newblock Angew. Chem. Int. Ed. {\bf 53}, 3201 (2014).

\bibitem{kagan10a}
D.~Kagan et~al.,
\newblock Small {\bf 6}, 2741 (2010).

\bibitem{sundararaj10a}
S.~Sundararajan, S.~Sengupta, M.~E. Ibele, and A.~Sen,
\newblock Small {\bf 6}, 1479 (2010).

\bibitem{gao12a}
W.~Gao et~al.,
\newblock Small {\bf 8}, 460 (2012).

\bibitem{gao14a}
W.~Gao and J.~Wang,
\newblock Nanoscale {\bf 6}, 10486 (2014).

\bibitem{lien99a}
H.-L. Lien and W.-x. Zhang,
\newblock J. Environ. Eng. {\bf 125}, 1042 (1999).

\bibitem{kim04a}
M.~J. Kim and K.~S. Breuer,
\newblock Phys. Fluids {\bf 16}, L78 (2004).

\bibitem{hernandez05a}
J.~P. Hernandez-Ortiz, C.~G. Stoltz, and M.~D. Graham,
\newblock Phys. Rev. Lett. {\bf 95}, 204501 (2005).

\bibitem{kim07a}
M.~J. Kim and K.~S. Breuer,
\newblock Anal. Chem. {\bf 79}, 955 (2007).

\bibitem{pushkin13}
D.~Pushkin and J.~Yeomans,
\newblock Phys. Rev. Lett. {\bf 111}, 188101 (2013).

\bibitem{cates12a}
M.~Cates,
\newblock Rep. Prog. Phys. {\bf 75}, 042601 (2012).

\bibitem{cates15a}
M.~Cates and J.~Tailleur,
\newblock Ann. Rev. Cond. Mat. Phys. {\bf 6}, 219 (2015).

\bibitem{brown14a}
A.~Brown and W.~Poon,
\newblock Soft Matter {\bf 10}, 4016 (2014).

\bibitem{ebbens12a}
S.~Ebbens, M.-H. Tu, J.~R. Howse, and R.~Golestanian,
\newblock Phys. Rev. E {\bf 85}, 020401 (2012).

\bibitem{ebbens14a}
S.~Ebbens et~al.,
\newblock EPL {\bf 106}, 58003 (2014).

\bibitem{howse07a}
J.~R. Howse et~al.,
\newblock Phys. Rev. Lett. {\bf 99}, 048102 (2007).

\bibitem{lee14a}
T.-C. Lee et~al.,
\newblock Nano Lett. {\bf 14}, 2407 (2014).

\bibitem{paxton04a}
W.~F. Paxton et~al.,
\newblock J. Am. Chem. Soc. {\bf 126}, 13424 (2004).

\bibitem{simmchen14a}
J.~Simmchen et~al.,
\newblock RSC Adv. {\bf 4}, 20334 (2014).

\bibitem{valadares10a}
L.~F. Valadares et~al.,
\newblock Small {\bf 6}, 565 (2010).

\bibitem{wang06a}
Y.~Wang et~al.,
\newblock Langmuir {\bf 22}, 10451 (2006).

\bibitem{moran10a}
J.~Moran, P.~Wheat, and J.~Posner,
\newblock Phys. Rev. E {\bf 81}, 065302 (2010).

\bibitem{sabass12b}
B.~Sabass and U.~Seifert,
\newblock J. Chem. Phys. {\bf 136}, 214507 (2012).

\bibitem{brown15a-pre}
A.~Brown, W.~Poon, C.~Holm, and J.~de~Graaf,
\newblock arXiv {\bf 1512.01778}, 1 (2015).

\bibitem{gibbs10a}
J.~Gibbs, N.~Fragnito, and Y.~Zhao,
\newblock Appl. Phys. Lett. {\bf 97}, 253107 (2010).

\bibitem{paxton05a}
W.~Paxton, A.~Sen, and T.~Mallouk,
\newblock Chem. Eur. J. {\bf 11}, 6462 (2005).

\bibitem{sabass10a}
B.~Sabass and U.~Seifert,
\newblock Phys. Rev. Lett. {\bf 105}, 218103 (2010).

\bibitem{sabass12a}
B.~Sabass and U.~Seifert,
\newblock J. Chem. Phys. {\bf 136}, 064508 (2012).

\bibitem{wang13b}
W.~Wang, T.-Y. Chiang, D.~Velegol, and T.~Mallouk,
\newblock J. Am. Chem. Soc. {\bf 135}, 10557 (2013).

\bibitem{popescu10a}
M.~Popescu, S.~Dietrich, M.~Tasinkevych, and J.~Ralston,
\newblock Eur. Phys. J. E {\bf 31}, 351 (2010).

\bibitem{golestanian07a}
R.~Golestanian, T.~B. Liverpool, and A.~Ajdari,
\newblock New J. Phys. {\bf 9}, 126 (2007).

\bibitem{buttinoni13a}
I.~Buttinoni et~al.,
\newblock Phys. Rev. Lett. {\bf 110}, 238301 (2013).

\bibitem{anderson89a}
J.~L. Anderson,
\newblock Annu. Rev. Fluid Mech. {\bf 21}, 61 (89).

\bibitem{brady11a}
J.~F. Brady,
\newblock J. Fluid Mech. {\bf 667}, 216 (2011).

\bibitem{sharifimood13a}
N.~Sharifi-Mood, J.~Koplik, and C.~Maldarelli,
\newblock Phys. Fluids {\bf 25}, 012001 (2013).

\bibitem{shklyaev14a}
S.~Shklyaev, J.~Brady, and U.~C{\'o}rdova-Figueroa,
\newblock J. Fluid Mech. {\bf 748}, 488 (2014).

\bibitem{degraaf15a}
J.~de~Graaf, G.~Rempfer, and C.~Holm,
\newblock IEEE Trans. NanoBiosci. {\bf 14}, 272 (2015).

\bibitem{michelin15a}
S.~Michelin and E.~Lauga,
\newblock Eur. Phys. J. E {\bf 38}, 1 (2015).

\bibitem{han96a}
P.~Han and D.~M. Bartels,
\newblock J. Phys. Chem. {\bf 100}, 5597 (1996).

\bibitem{sridhar83a}
T.~Sridhar and O.~Potter,
\newblock Chem. Eng. Commun. {\bf 21}, 47 (1983).

\bibitem{haynes13a}
W.~M. Haynes, editor,
\newblock {\em CRC handbook of chemistry and physics},
\newblock CRC press, Boca Raton, U.S.A., 93rd edition, 2013.

\bibitem{dunkel10a}
J.~Dunkel, V.~Putz, I.~Zaid, and J.~Yeomans,
\newblock Soft Matter {\bf 6}, 4268 (2010).

\bibitem{baraban12a}
L.~Baraban et~al.,
\newblock ACS Nano {\bf 6}, 3383 (2012).

\bibitem{kreuter13a}
C.~Kreuter, U.~Siems, P.~Nielaba, P.~Leiderer, and A.~Erbe,
\newblock Eur. Phys. J. ST {\bf 222}, 2923 (2013).

\end{thebibliography}
\end{document}